\documentclass[preprint,showpacs,preprintnumbers,amsmath,amssymb,nofootinbib]{revtex4}

\usepackage{etex}

\usepackage{amsthm,amscd,amsbsy,array}
\usepackage{bm}
\usepackage{soul} 

\usepackage{graphics,graphicx,xcolor}

\usepackage{soul} 

\usepackage[colorlinks=true, pdfstartview=FitV, linkcolor=blue, citecolor=blue, urlcolor=blue]{hyperref} 

\newcommand{\cyan}{\textcolor{cyan}}
\newcommand{\magenta}{\textcolor{magenta}}
\newcommand{\red}{\textcolor{red}}
\newcommand{\blue}{\textcolor{blue}}

\newcommand{\gb}{\colorbox{green}}

\newenvironment{redtext}{\color{red}}{\ignorespacesafterend}
\newenvironment{bluetext}{\color{blue}}{\ignorespacesafterend}

\newcommand{\bblue}{\begin{bluetext}}
\newcommand{\eblue}{\end{bluetext}}
\newcommand{\bred}{\begin{redtext}}
\newcommand{\ered}{\end{redtext}}

\numberwithin{equation}{section}

\let\ssection=\section
\renewcommand{\section}{\setcounter{equation}{0}\ssection}

\def\besub{\begin{subequations}}
\def\balign{\begin{align}}
\def\esub{\end{subequations}}
\def\ealign{\end{align}}

\newcommand{\ba}{{\bf a}}

\newcommand{\cA}{{\mathcal{A}_{+}}}
\newcommand{\cB}{{\mathcal{A}_{\times}}}

\newcommand{\diag}{\mathrm{diag}}

\newcommand{\bbR}{\mathbb{R}}

\renewcommand{\Re}{\mathrm{Re}}

\newcommand{\bX}{{\bf X}}

\newcommand{\bY}{{\bf Y}}

\def\smallover#1/#2{\hbox{$\textstyle\frac{#1}{#2}$}} %

\def\parag{\hfil\break} 
\def\kikezd{\parag\underbar}

\def\benu{\begin{enumerate}}
\def\eenu{\end{enumerate}}
\def\beq{\begin{equation}}
\def\eeq{\end{equation}}
\def\beqa{\begin{eqnarray}}
\def\eeqa{\end{eqnarray}}
\def\nn{\nonumber}
\def\barray{\left(\begin{array}}
\def\earray{\end{array}\right)}
\def\barraynb{\begin{array}}
\def\earraynb{\end{array}}

\def\?{\quad{\gb{\fbox{\texttt{?}}\;}}\quad}
\def\p{{\partial}}

\def\v0{\mathbf{0}}

\def\beq{\begin{equation}}
\def\eeq{\end{equation}}
\def\bea{\begin{eqnarray}}
\def\eea{\end{eqnarray}}

\def\p{\partial}

\def \p{{\partial}}

\def\6{\partial}
\def\7{\tilde}
\def\8{\widehat}

\newcommand{\const}{\mathop{\rm const.}\nolimits}
\newcommand{\half }{\frac{1}{2}}

\def\smallover#1/#2{\hbox{$\textstyle\frac{#1}{#2}$}} %
\def\smallcirc{{\raise 0.5pt \hbox{$\scriptstyle\circ$}}}
\def\2{{\smallover1/2}}

\let\ssection=\section
\renewcommand{\section}{\setcounter{equation}{0}\ssection}

\begin{document} 
\preprint{[arXiv:1802.09061v3 [gr-qc]].}

\title{Velocity Memory Effect for Polarized
 Gravitational Waves
\\[6pt]}

\author{
P.-M. Zhang${}^{1}$\footnote{e-mail:zhpm@impcas.ac.cn},
C. Duval${}^{2}$\footnote{
mailto:duval@cpt.univ-mrs.fr},
G.W. Gibbons$^{3}$\footnote{
e-mail: G.W.Gibbons@damtp.cam.ac.uk},
P. A. Horvathy${}^{1,4}$\footnote{mailto:horvathy@lmpt.univ-tours.fr},
}

\affiliation{
$^1$Institute of Modern Physics, Chinese Academy of Sciences, Lanzhou, China
\\
$^2$Aix Marseille Univ, Universit\'e de Toulon, CNRS, CPT, Marseille, France
\\
$^3${\small D.A.M.T.P., Cambridge University, U.K.}
\\
$^4$ Laboratoire de Math\'ematiques et de Physique
Th\'eorique, Universit\'e de Tours, France
}

\date{\today}


\begin{abstract} Circularly polarized  gravitational sandwich waves exhibit, as do their linearly polarized counterparts, the Velocity Memory Effect: freely falling test particles in the flat after-zone fly apart along straight lines with constant velocity.
In the inside zone their trajectories combine oscillatory  and rotational motions in  a complicated way. For circularly polarized periodic gravitational waves some trajectories remain bounded, while others spiral outward. These waves admit an additional ``screw'' isometry beyond the usual five. The consequences of this extra symmetry are explored.
\\[8pt]
{JCAP (to be published) }
\end{abstract}

\maketitle

\tableofcontents

\section{Introduction}\label{Intro}

In the \emph{Velocity Memory Effect test particles initially at rest fly apart with nonvanishing constant velocity after a  burst of a gravitational wave (GW) has passed} \cite{Ehlers,Sou73,GriPol,BraTho,BoPi89,ShortMemory,LongMemory, ImpMemory,Lasenby}. This statement (which contradicts that of Zel'dovich and Polnarev \cite{ZelPol}, who claimed simple displacement with zero relative velocity), appears, in a clear and precise form in a seminal paper of Ehlers and Kund \cite{Ehlers} published  11 years before \cite{Sou73}, 12 years before \cite{ZelPol} and 23 years before \cite{BraGri}.  In the authors' own words~:
\textit{\narrower ``After a pulse has swept over the particles they have constant {\it velocities}\dots ''}.
  The effect, although  very small, could theoretically be observed through the Doppler effect \cite{BraTho,Lasenby}. For further details and references see, e.g., \cite{Fava,Harte12,Winicour14,Madler16,Lasky16}.

Our previous investigations \cite{ShortMemory,LongMemory, ImpMemory}  focused at linearly polarized sandwich waves \footnote{The spacetime of a sandwich wave is flat outside an  interval $[U_i,U_f]$ of a ``non-relativistic time'' coordinate, $U$.}.

However those recent observations by LIGO, and VIRGO \cite{LIGO1,LIGO2} which  revolutionized the field  indicate that these
gravitational waves are \emph{not} linearly polarised, and one may wonder if our findings pertain to linear polarisation.
\goodbreak

In this paper we extend our investigations to polarized gravitational waves  with the conclusion that they behave
essentially as linearly polarised do, albeit  with considerable additional complications.

We start by studying  \emph{circularly polarized (approximate) sandwich waves}. Circular polarization
is indeed expected to arise  for a binary (either black holes or neutron stars) and looking along the orbital rotation axis, this is circularly polarized. For core-collapse supernov{\ae} if the star was rotating beforehand, then circular polarization will be induced by this component. The other aspect is that for rotating systems the rotation axis is the one along which the GW wave emission is generally greatest, on average one will preferentially see circularly polarized emission. Circularly polarized sandwich waves might arise, e.g., for coalescing  black holes \cite{LIGO1} or neutron star merger \cite{LIGO2}.
Our results may therefore apply to the suggestion of \cite{Moore:2017ity}  that  astrometric data  from GAIA may allow detection of long wavelength gravitational radiation emitted by binary systems. See also \cite{Klioner:2017asb}.

We  begin by  studying  what happens for a simplified model, namely for an \emph{oscillating profile combined with a Gaussian envelope}, see our eqns.
(\ref{genBrink}--\ref{polgaussprof}) depicted in Fig.\ref{AB} below, which could be thought of as (a naive) idealization of those  ``chirps'' in Fig.1 of \cite{LIGO1} \footnote{Our naive plots  actually fit better neutron star merger, cf. Fig.2  of \cite{LIGO2}.}.

Although we did not succeed  finding  analytic solutions let alone for our simplified model, our numerical calculations indicate that, while the motion in the inside-zone $[U_i,U_f]$ is highly complicated, in the (approximately) flat before- and after-zones of a(n approximate) sandwich wave,
 \emph{the motion is still along straight lines with constant velocity}, consistent with \dots \emph{Newton's First law}.

Then, dropping the Gaussian envelope, we turn to \emph{purely periodic} (non-sandwich) gravitational wave trains,
which are suspected to arise (but unsuccessfully searched for so far) from
rapidly rotating neutron stars \cite{LIGOPer},
and/or for primordial waves in the inflationary universe \cite{Kamion}.

Periodic waves, modelled by the uniformly rotating profile in (\ref{PerProf}) have interesting theoretical properties. In the linearly polarized case analytic solutions can be found in terms of Mathieu functions.
Another remarkable property is that they carry, in addition to the general 5-parameter isometry \cite{BoPiRo,Sou73,exactsol,Sippel,Torre, Carroll4GW} an intriguing ``screw symmetry'' \cite{exactsol, Sippel,Carroll4GW}.
\goodbreak

\section{Motions in a plane Gravitational Wave}
\label{Traject}

In Brinkmann (B) coordinates $(\bX,U,V)$
the profile of a plane GW is given by
 the symmetric and traceless $2\times2$ matrix
$H(U)=H_{ij}(U)$ \cite{Brink,BoPiRo},
\begin{subequations}
\begin{align}
ds^2=\delta_{ij} dX^i dX^j + 2 dU dV + H_{ij}(U) X^i X^j dU^2,
\label{Bmetric}
\\[6pt]
H_{ij}(U){X^i}{X^j}=
\half{\cA}(U)\big((X^1)^2-(X^2)^2\big)+\cB(U)\,X^1X^2,
\label{Bprofile}
\end{align}
\label{genBrink}
\end{subequations}
where $\cA$ and $\cB$ are the amplitudes of the $+$ and $\times$ polarization states.
Let us record for further reference that our 4D GW can be viewed as the ``Bargmann manifold'' of a $2+1$ dimensional non-relativistic system, namely of a time-dependent anisotropic oscillator  \cite{Eisenhart,Bargmann}.
The motion of a spinless test particle in the (\ref{genBrink}) background is described by geodesics,
$X^\mu(\sigma)$,
whose equations of motion
can be  derived by varying the natural geodesic Lagrangian
$\half g_{\mu\nu}\dot{X}^\mu\dot{X}^\nu$, i.e.,
\beq
L= \half \bigl(\dot X_i\dot X^i + 2\dot U \dot V +
H_{ij}(U)X^i X^j\, (\dot U)^2\bigr)
\label{geoLag}
\eeq
where $\dot{X}^\mu = {d X^\mu}/{d\sigma}$.
Variation with respect to $V$ yields that
$
\ddot U = 0
$
allowing us to choose $\sigma=U$ and providing us finally with
\begin{subequations}
\begin{align}
& \dfrac{d^2\bX}{dU^2} - H(U)
\,\bX = 0,
\qquad\qquad
 H(U)= \half\barray{lr}
{\cA} &{\cB}\\{\cB} & -{\cA}\earray,
\label{ABXeq}
\\[12pt]
& \dfrac {d^2 V}{dU^2} + \dfrac{1}{4} \dfrac{d{\cA}}{dU\,}\left((X^1)^2 - (X^2)^2 \right )
+{\cA}\left(X^1 \dfrac{dX^1}{dU\,} - X^2\dfrac{dX^2}{dU\,} \right )
\nn
\\[8pt]
&\qquad\,+\dfrac{1}{2} \dfrac {d{\cB}}{dU\,} X^1 X^2
+ {\cB}\left(X^2\dfrac{dX^1}{dU\,} + X^1\dfrac{dX^2}{dU\,} \right)
= 0.
\label{ABVeq}
\end{align}
\label{ABeqs}
\end{subequations}\vskip-5mm

 In what follows, we limit our investigations (as before \cite{ShortMemory,LongMemory,ImpMemory}) to the motion of  particles which are at rest in the before-zone, $\bX(U)=\bX_0$
for $U\leq U_i$.

\section{Circularly Polarized Waves}

If ${\cA}(U) = 0 $ or ${\cB}(U) = 0$,
  the wave is linearly polarized; the motions were studied in  our previous papers \cite{ShortMemory,LongMemory,ImpMemory}.
Turning  to {polarized waves}
and approximating the sandwich by a Gaussian, we consider now
\begin{subequations}\vspace{-2mm}
\begin{align}
{\cA} (U) &= A_0\,\frac{\lambda}{\sqrt{\pi}}\,e^{-\lambda^2U^2} \cos(\omega U),
\\
{\cB} (U) &= B_0\,\frac{\lambda}{\sqrt{\pi}}\,e^{-\lambda^2U^2}
\cos(\omega U-\phi).
\end{align}
\label{polgaussprof}
\end{subequations}
 For simplicity we chose, henceforth, $A_0=B_0=C^2=\const$ and
 $\phi=\pi/2$ \footnote{Changing the relative phaseshift $\phi$ has a strong influence on geodesics. For $\phi=0$ we have ${\cA} \propto {\cB}$ and the wave is once again linearly polarized.}.
 Then  the polarization is circular with an $U$-dependent amplitude, $D(U)= C^2({\lambda}/{\sqrt{\pi}})\,\exp[{-\lambda^2U^2}]$ \footnote{More sophisticated formul{\ae} were considered in  \cite{Blanchet}:
\beq
A_+=A\big(\omega(U)\big)^{2/3}\cos\varphi(U), \quad
\varphi(U)=\int_{U_0}^U\omega(U')dU', \quad
\omega(U)=\big(k(U-U_0)\big)^{3/2}\; .\;
\label{Blanchetprof}
\eeq
}\,,
\beq
\cA =
D(U)\, \cos(\omega U),
\qquad
{\cB}=
D(U)\,\sin(\omega U).
\label{pol}
\eeq

Our main interest lies in determining $\bX(U)$; displacements of the $V$ coordinate are of the second order \cite{LongMemory}. Eqn. (\ref{ABXeq}) describes, in ``Bargmann'' language, an anisotropic harmonic oscillator in transverse space with time-dependent (rotating) force.
 The latter is non-vanishing only in the (approximate) inside zone
\footnote{The ``before'', $U < U_i$, ``inside'' $U_i \leq U \leq U_f$ and ``after'' $U_f < U$ zones are defined only approximately by the requirement that the Gaussian factor be small.}.

\begin{figure}[h]
\hskip-3mm
\includegraphics[scale=.29]{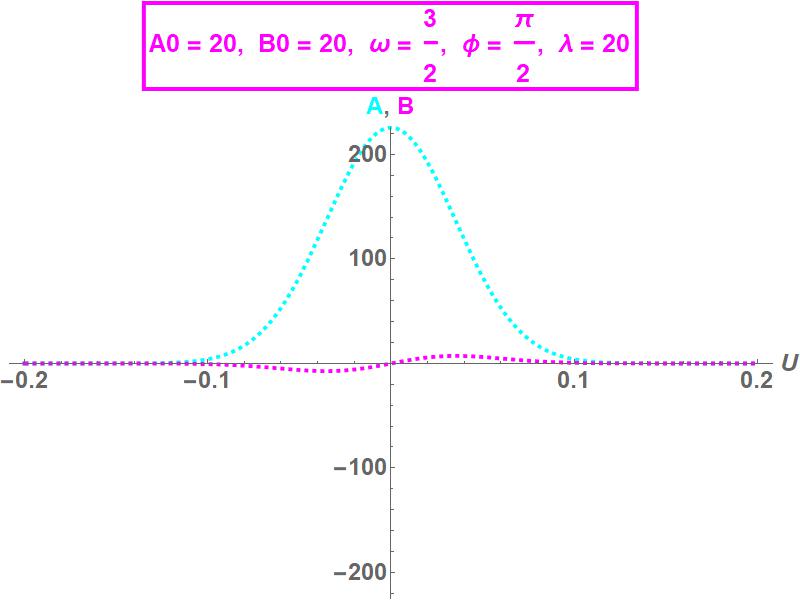}\,
\includegraphics[scale=.29]{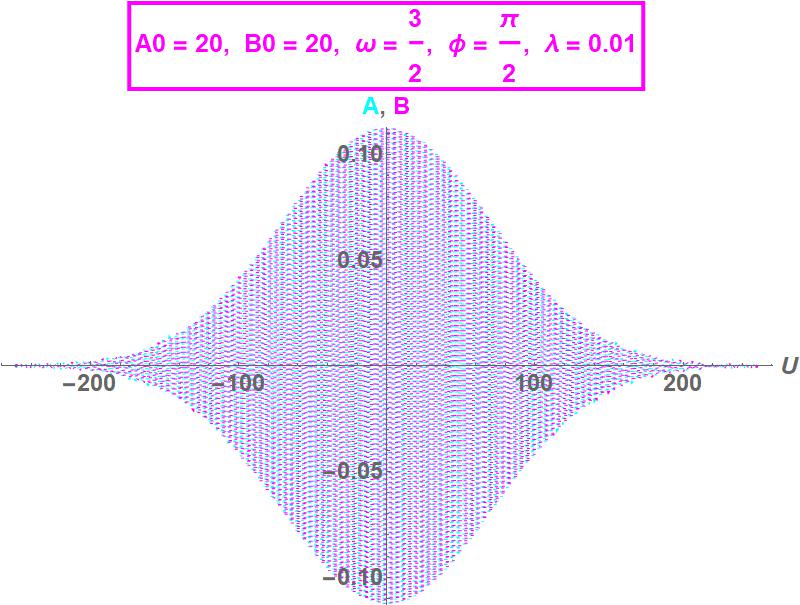}
\\
\hskip-3mm \small{(i)}\hskip78mm \small{(ii)}\\
\vskip-3.5mm
\caption{\textit{\small Profile of circularly polarized sandwich waves for (i) large and (ii) small thickness parameter $\lambda$.
For $\lambda\to\infty$ the Gaussian profile approximates
a polarized  impulsive wave, cf. (\ref{impPol}), and for $\lambda\to0$ it becomes weak but wide. The colors \cyan{\bf{cyan}} and \magenta{\bf{magenta}} refer to the orthogonal components $\cyan{\cA}$ and $\magenta{\cB}$ of the spin-$2$ field $H(U)$. Note the different scales of the figures.}}
\label{AB}
\end{figure}

\begin{figure}[h]
\hskip-2mm
\includegraphics[scale=.28]{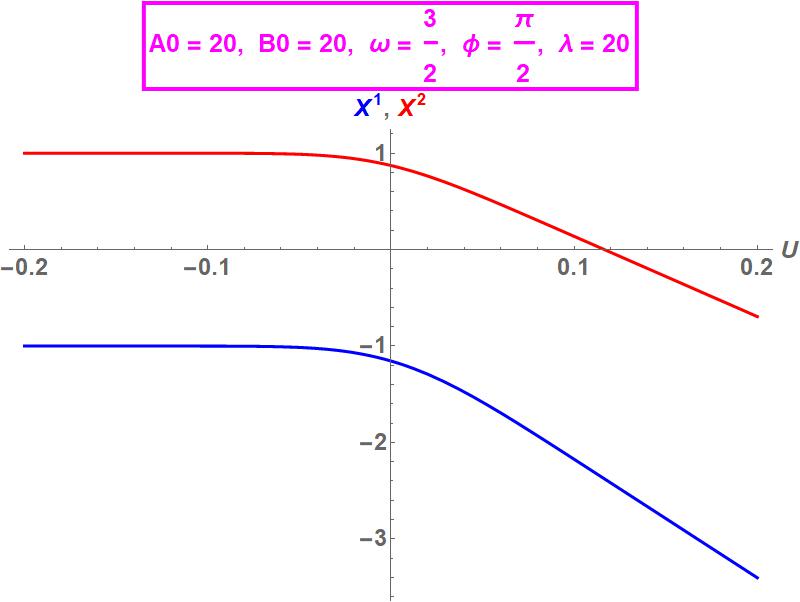}\;
\includegraphics[scale=.3]{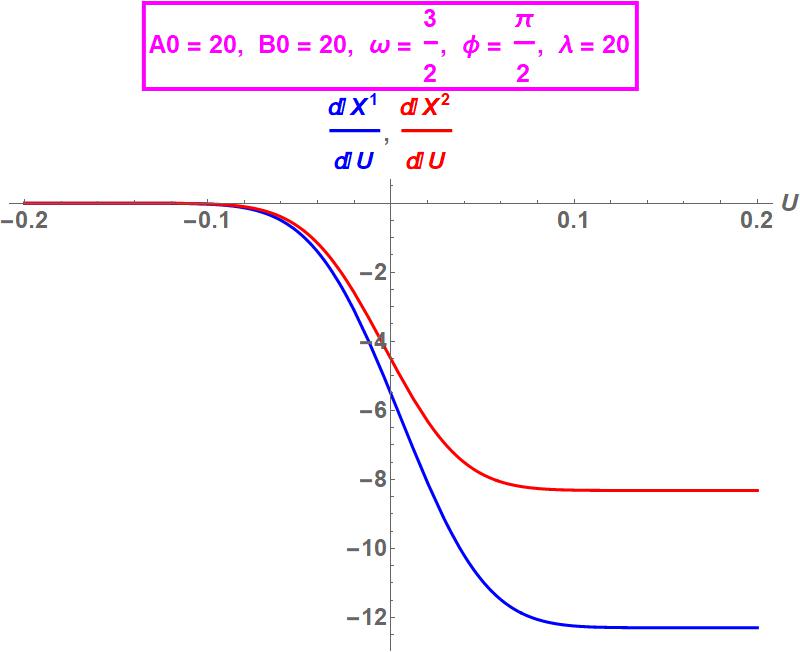}
\\
\caption{\textit{\small 
Trajectories 
and velocities
after the passage of a circularly polarized Gaussian sandwich wave (\ref{polgaussprof}) in
the impulsive limit $\lambda\to\infty$. The \blue{\bf blue} and \red{\bf red} colors refer to the transverse components $\blue{X^1}$ and $\red{X^2}$.
The trajectory is bent in the (narrow) inside zone and then escapes with non-vanishing constant velocity in the flat after-zone.
}}
\label{largeltrajvel}
\end{figure}

\begin{figure}
\hskip-4mm
\includegraphics[scale=.28]{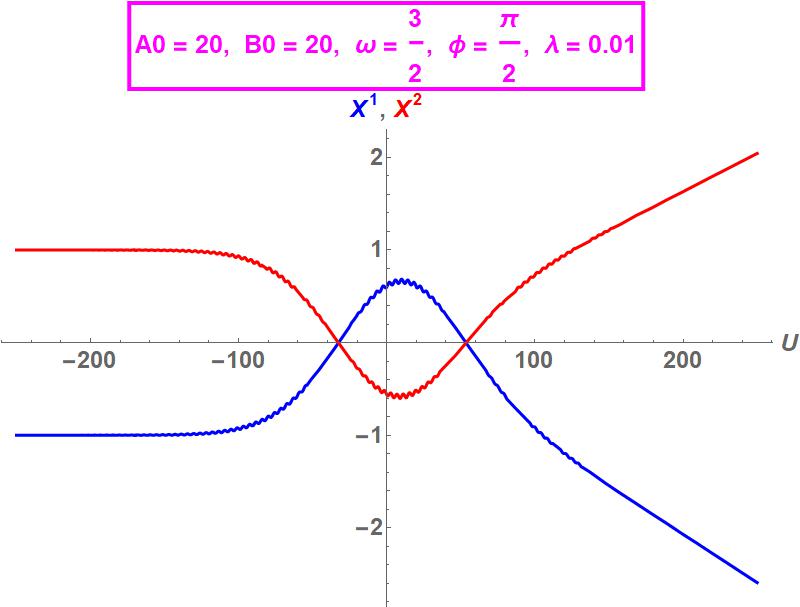} 
\includegraphics[scale=.31]{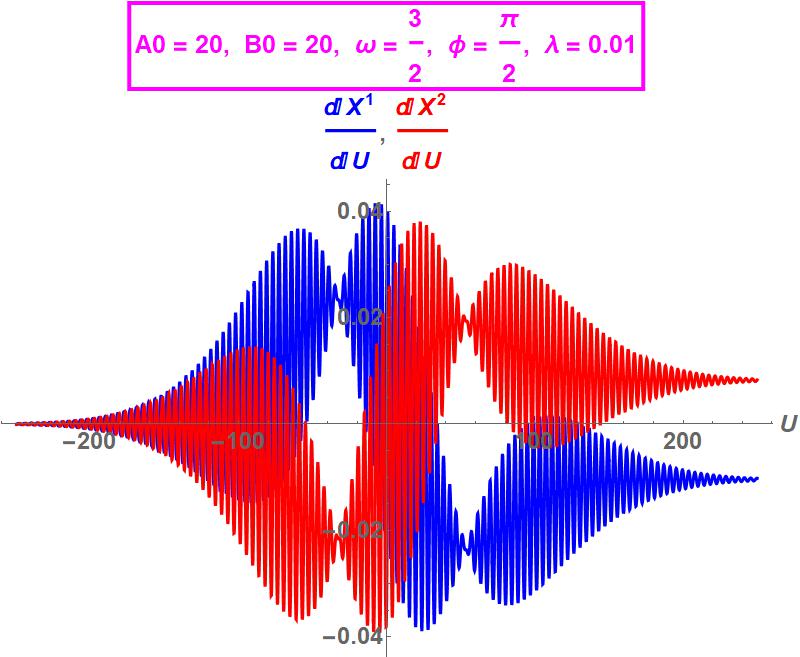}
 \vskip-3mm
\caption{\textit{\small  
Trajectories 
and velocities
for small $\lambda$, describing test particle motion in a wide but weak circularly polarized Gaussian sandwich GW (\ref{polgaussprof}).  The \blue{\bf blue} and \red{\bf red} colors refer to the transverse components $\blue{X^1}$ and $\red{X^2}$. A particle initially at rest has a complicated motion in the (approximate) inside zone however it escapes with non-zero constant velocity in the flat  after-zone.
}}
\label{smallltrajvel}
\end{figure}

The profiles for different width-parameters $\lambda$ are depicted in Fig.\ref{AB}.
The numerically calculated trajectories and velocities,
 shown in Figs.\ref{largeltrajvel} and \ref{smallltrajvel}, hint at the following behaviour:

\benu
\item
In the \emph{large-$\lambda$} regime (i) the Gaussian is thin and high and the motion is roughly along straight lines with constant velocity, except in a short ($\sim O(1/\lambda)$ transitory ``inside'' region, where the trajectory is sharply bent and the velocity changes rapidly from zero to some non-zero value Fig.\ref{largeltrajvel} --- reminiscent of motion in an impulsive wave \cite{Penrose,BarHog,PodSB,ImpMemory}.

\item
In the \emph{small-$\lambda$} regime (ii) the profile is wide and low, with many oscillations in the inside zone. Apart of some fine ``denting'', the motion is ``reasonably regular'' in the inside zone, see Fig.\ref{smallltrajvel}. The effect of  ``denting'' becomes important, though, when the velocity is plotted. The trajectories suffer a rather weak rotation.
\eenu

In the outside zones, $U < U_i$ and $U > U_f$, everything smooths out for both regimes:  the trajectory appears, just like in the linearly polarized case, to follow straight lines with constant velocity.
The difference between the large and small-$\lambda$ regimes lies in the inside zone: motion in the after zone is always simple.

Although we have not been able to find analytic solutions, we have arguments to confirm our observations made above.

Firstly, the behaviour in the outside zones is obvious from the Bargmann point of view, which says that the  non-relativistic  motions in $d+1$ dimensions are the projections of lightlike geodesics in a $(d+1,1)$ dimensional Lorentz manifold along a covariantly constant null direction \cite{Eisenhart, Bargmann}. Our metric (\ref{genBrink}) describes, in particular, non-relativistic motions in 2+1 dimensions in the potential $\half{} H_{ij}X^iX^j$.
Where $H\approx 0$, we have therefore a \emph{free non-relativistic particle}. Our observation made above is therefore \dots \emph{Newton's First Law}.

Next, integrating (\ref{ABXeq}) from $U_i$ to $U_f$ yields the \emph{accumulated change of  velocity} as
\beq
\Delta\dot{\bX}=
\int_{U_i}^{U_f}\!\! H(U)\,\bX(U) dU\,
=
\barray{l}
\half\displaystyle\int_{U_i}^{U_f}\!\!
{\small \big({\cA}(U)X^1(U)+\cB(u)X^2(U)\big)\, dU}
\\[14pt]
\half\displaystyle\int_{U_i}^{U_f}\!\!
{\small\big(
{\cB}(U)X^1(U)-\cA(u)X^2(U)\big)\, dU}\,,
\earray
\label{genveljump}
\eeq
which generalizes  eqn (VI.1) of \cite{ImpMemory} from linear to circular polarization. The integral can be evaluated numerically and the result is \emph{consistent} with what is seen on the figures,
both in  the large- and small-$\lambda$ regimes,
Figs.\ref{largeltrajvel} and  \ref{smallltrajvel}, respectively. We checked this for $\lambda = 20$ by evaluating the integral numerically.

When $\lambda \to \infty$  the profile (\ref{genBrink})-(\ref{polgaussprof}) goes over to the impulsive form \cite{Penrose,BarHog,PodSB,ImpMemory}
\begin{equation}
\cA(U) = C\, \delta (U) \cos(\omega U),
\qquad
\cB(U)= C\,\delta(U) \sin(\omega U).
\label{impPol}
\end{equation}
The remarkable result is that evaluating the velocity jump (\ref{genveljump}) the off-diagonal component $\cB$, and in fact the circular frequency $\omega$ \emph{drop out}, leaving us with the \emph{same expression as in the linearly polarized case} \cite{ImpMemory}, namely with \footnote{
For arbitrary
$A_0, B_0$ and $\phi$ (\ref{c0imp}) is generalized to
\beq
\dot{\bX} (0^+) = c_0 {\bX}_0
\quad\text{where}\quad
c_0 = \half \barray{cc}
A_0&B_0\cos\phi
\\
B_0\cos\phi &-A_0\earray.
\label{c0impgen}
\eeq}
\begin{equation}
\dot {\mathbf X} (0^+) = c_0 {\mathbf X}_0
\quad\text{where}\quad
c_0 = \half C~ \diag(1,-1).
\label{c0imp}
\end{equation}

\emph{At the impulsive limit the polarization has no effect}, consistent with the findings of Podolsk\'y and Vesel\'y  \cite{PodVes}, who argued that the resulting particle motions are identical regardless of the
 sandwich  we started with before shrinking. An intuitive explanation is that when $\lambda\to\infty$ the inside zone shrinks to a single instant $U\!=\!0$ and the extra structure has no time to act.

\begin{figure}[h]
\begin{center}
\includegraphics[scale=.48]{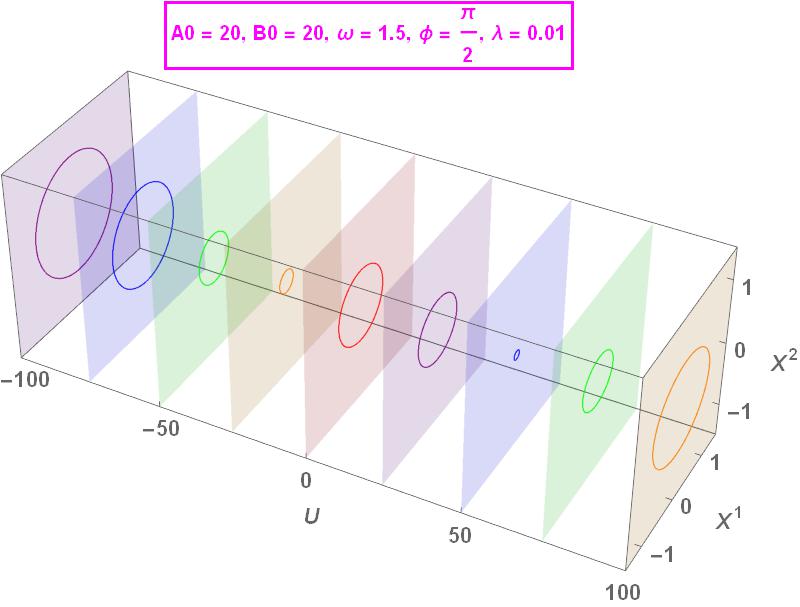}
\end{center}\vskip-9mm
\caption{\textit{\small ``Tissot'' diagram~:
the trajectories in the inside zone of a circularly polarized gravitational wave with wide ($\lambda=.01$) Gaussian profile ``breathe", as they do for linear polarisation.}}
\label{polgaussTissot}

\end{figure}

\begin{figure}[h]
\begin{center}
\includegraphics[scale=.28]{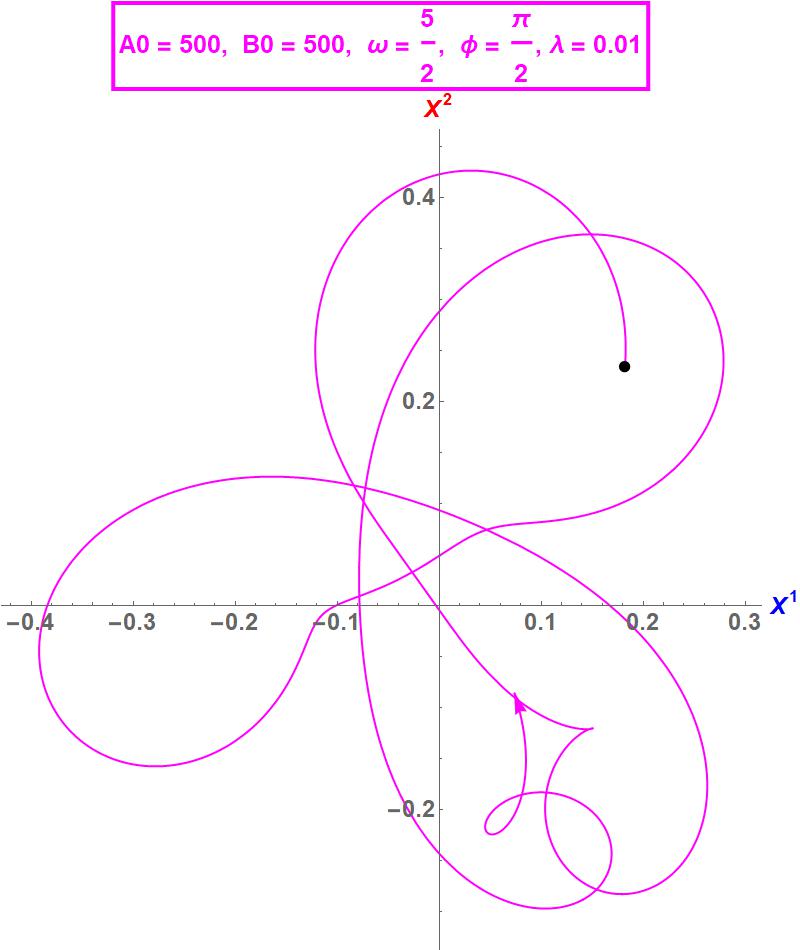}
\end{center}\vskip-9mm
\caption{\textit{\small In the inside zone of a circularly polarized GW the transverse trajectory precesses in a rather complicated way.}}
\label{smalllambdaprec}
\end{figure}

The ``Tissot'' diagram (Fig.\ref{polgaussTissot}), which shows the evolution of a small circle of particles initially at rest, exhibits the characteristic ``breathing'', seen before for linear polarization \cite{ShortMemory,LongMemory,ImpMemory}.
Polarized waves exhibit also a \emph{new effect}~: unlike in the linearly polarized case, the ``breathing" is combined with a rather complicated precession in transverse space, shown in
Fig.\ref{smalllambdaprec}.


\section{Periodic waves}

Waves with periodic  profile
\beq
{\cA} (U) = A_0 \cos(\omega U),
\qquad
{\cB} (U) = B_0\,
\cos(\omega U-\phi),
\label{PerProf}
\eeq
(although not yet observed) are believed to be generated by, e.g., rapidly rotating neutron stars \cite{LIGOPer}, or in the early inflationary universe \cite{Kamion}. They differ from our previously considered sandwich waves by the absence of damping.  Such wave behave oppositely to  impulsive waves: intuitively, they have only an inside zone.

\begin{figure}[h]
\begin{center}
\includegraphics[scale=.42]{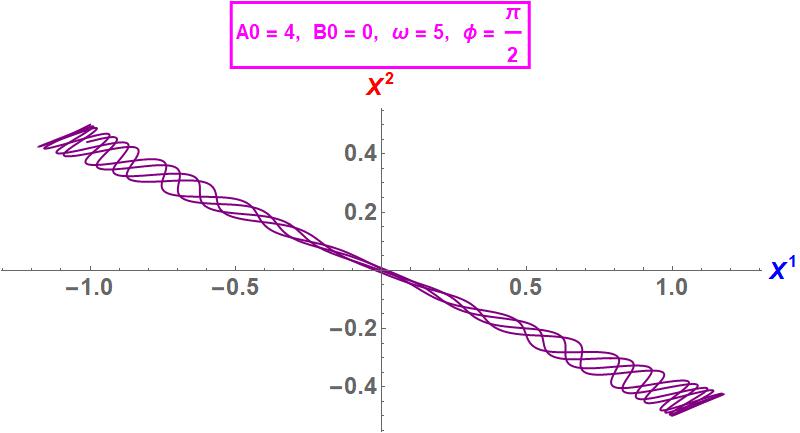}\\
\end{center}\vskip-9mm
\caption{\textit{\small
In a linearly polarized periodic wave with no relative phase shift, $\phi=0$, the transverse coordinate $\bX(U)$ oscillates along a ``dented" straight line.
The initial conditions are
$\dot X^1 (U\!=\!0) = \dot X^2 (U\!=\!0) = 0$ (at rest for $U\!=\!0$) at initial position $X^1(U\!=\!0)= -1,\, X^2(U\!=\!0) = 1/2 $.}}
\label{linpolper}
\end{figure}

We first note that in the linearly polarized case $B_0=0$  the matrix $H(U)$ in (\ref{ABXeq}) is diagonal, and
the equations of motion can be solved analytically : they separate, leaving us with two uncoupled Mathieu equations,
\beq
\ddot \bX - \half \cos (\omega U) \barray{cc}
A_0,&0
\\
0 &-A_0
\earray
\bX= 0,
\label{Apolpertraj}
\eeq
whose general solutions are combinations of
  even and odd (Mathieu cosine resp. Mathieu sine) functions $C(a,q,x)$ resp. $S(a,q,x)$.
Choosing the initial conditions
$\dot \bX(U\!=\!0)= 0$ (particle at rest at $U\!=\!0$),
the solutions are pure
Mathieu cosine functions,
\begin{subequations}
\begin{align}
X^1 (U) &= c_1\, C\big(0, A_0/ \omega^2, -(\omega/2)U\big),
\\
X^2 (U) &= c_3\, C\big(0, -A_0/ \omega^2, -(\omega/2) U\big),
\end{align}
\label{PartMathSol}
\end{subequations}
where the coefficients $c_1$ and $c_3 $ are determined by the initial conditions $\bX(0)$. Plotting either the Mathieu cosine or alternatively, solving the equations numerically, yields Fig.\ref{linpolper}.

In the skew-diagonal case $A_0=0$ the polarization is again linear, and the problem is equivalent to the previous one, since it can be brought to diagonal form by introducing new coordinates,
$Y^1 = (X^1 + X^2),\,  Y^2 = (X^1 - X^2)$,
yielding uncoupled Mathieu equations for the latter.
If the phase lag  $\phi=0$, we obtain again  (\ref{PartMathSol}).

\begin{figure}[h]
\begin{center} 
\includegraphics[scale=.42]{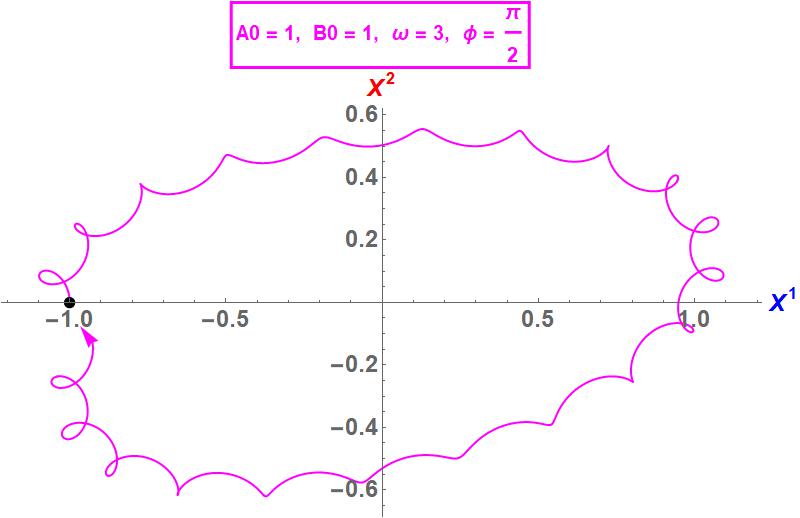}
\end{center}\vskip-9mm
\caption{\textit{\small The (transverse-space) trajectories in the periodic GW (\ref{PerProf}) are reminiscent of ``ellipses dented with epicycles".}}
\label{period-2d-traj}
\end{figure}

The general case is a combination of the previous ones, however we could  find numerical solutions only. Limiting ourselves again to circular polarization, $A_0=B_0=C^2$,
we found that the trajectory is a sort of
 ``ellipse, dented with an epicycle" cf. Fig.\ref{period-2d-traj} \footnote{Rotating motion was found also in a different background \cite{KuMa}.}.
 The choice of $\phi$ appears to have little effect .
 Choosing  $\phi=\pi/2$, for example, yields a ``dented elliptic'' trajectory along which the particle goes around: the trajectory is similar to the one in Fig.\ref{period-2d-traj}.

\begin{figure}[h]
\begin{center}
\includegraphics[scale=.48]{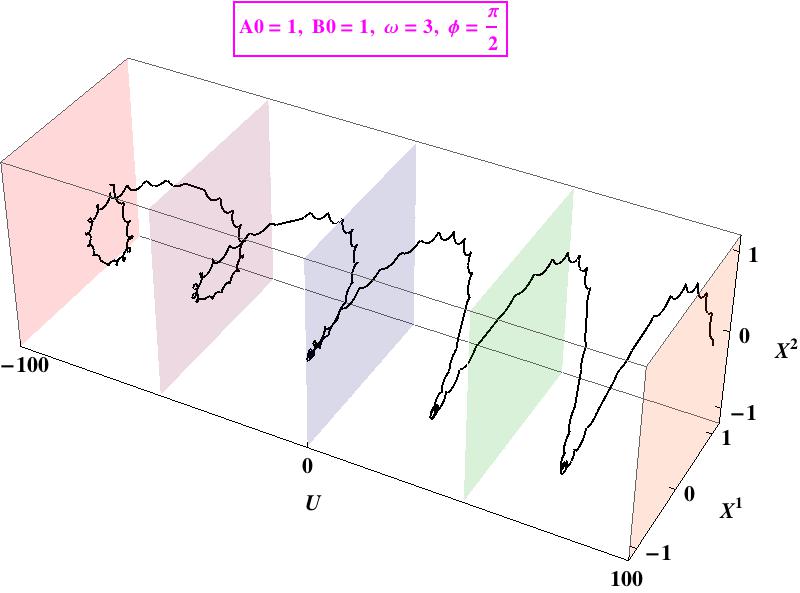}
\end{center}\vskip-8mm
\caption{\textit{\small For appropriate initial conditions the highly ``dented'' trajectories in the periodic GW background (\ref{PerProf}) spirals outward with a growing radius.}}
\label{Per-traj-Tissot}
\end{figure}

\section{Symmetries of circularly polarised periodic waves}

General plane waves have a 5-parameter isometry group  \cite{BoPiRo,exactsol,Sippel,Torre}, recently identified as the Carroll group with broken rotations \cite{Carroll4GW}~: for \blue{a} time-dependent anisotropic profile $U$-translational and transverse-space rotational symmetry are both broken.  However for a periodic wave, the two broken transformations can be combined to yield an additional, $6$th isometry \cite{exactsol,Sippel,Carroll4GW}. Let us consider indeed the periodic (Brinkmann) profile  in the particular case
$A_0=B_0=C^2, \, C>0$ and $\phi=\pi/2$ in (\ref{PerProf}), i.e.,
\begin{equation}
H_{ij}(U)X^iX^j
=
{C^2}\left(\frac{1}{2}\cos(\omega{}U)\,\left[(X^1)^2-(X^2)^2\right]+\sin(\omega{}U)\,X^1X^2\right)
\label{Kperiodic}
\end{equation}
where $\omega=\const$.
Using complex coordinates, $Z=X^1+iX^2$, (\ref{Kperiodic}) is
\beq
H_{ij}X^i X^j = \dfrac{C^2}{2}\, \Re (e^{-i\omega  U} Z^2 ) \,,
\label{Kcomplex}
\eeq
which is plainly invariant under the transformation
\beq
U\to U+e,\qquad
Z \to  e^{i\half\omega e} Z,\qquad
V\to{}V\,,
\label{compscrew}
\eeq
or in real form,
\begin{equation}
U\to{}U+e,
\qquad
\bX\to{}R_{\half\omega e}\bX,
\qquad
V\to{}V\,,
\label{screwtransf}
\end{equation}
where $R_{(\,\cdot\,)}$ is the rotation matrix in the plane.
These ``screw transformations'', which combine $U$-translations with transverse rotations, are therefore isometries for all $e\in\bbR$ \cite{exactsol,Sippel,Carroll4GW}; they belongs to the Bargmann group (itself a subgroup of the Poincar\'e group), but \emph{not} to its Carroll subgroup (for which $U$ is fixed).

A nice insight is due to P. Kosinski \cite{Kosinski} to whom we are grateful for allowing us to include the following. From the Bargmann point of view \cite{Eisenhart,Bargmann} the profile (\ref{Kperiodic})
describes an anisotropic oscillator in the transverse plane with time-dependent frequency. Viewed as a classical mechanical system, its motion can be described by the Lagrangian
\beq
L = \half \dot{\bX}^2 +\frac{C^2}{4}\Big(\cos\omega t\big((X^1)^2-(X^2)^2\big) + 2\sin\omega t\,X^1X^2
\Big),
\label{tdLag}
\eeq
for which both rotations and time translations are manifestly broken due to anisotropy and to time-dependence, respectively.
We record for further use the associated equations of motion,
\besub\vspace{-5mm}
\begin{align}
\ddot X^1 & = \dfrac{C^2}{2} \Big (X^1 \cos \omega t + X^2 \sin \omega t \;\Big)\,,
\\[4pt]
\ddot X^2 & =  \dfrac{C^2}{2} \Big(X^1 \sin \omega t  - X^2 \cos \omega t \;\Big)\,,
\end{align}
\label{Xeqmot}
\esub
which are coupled, driven by a time-dependent rotating linear force.
However, switching to a rotating frame by setting
\beq
\barray{c}X^1\\ X^2\earray
=
\barray{rcr}\cos\half \omega t &
&-\sin\half \omega t\\
\sin\half \omega t & &\cos\half \omega t\earray
\barray{c}Y^+\\ Y^-\earray,
\label{rotframe}
\eeq
the Lagrangian (\ref{tdLag}) can be written as
\besub
\begin{align}
&L = \half \dot{\bY}^2 + \dfrac{\omega}{2} \,\Big(\dot{Y}^-Y^+-
Y^-\dot{Y}^+\Big)+
\half \Omega_+^2\, (Y^+)^2+\half \Omega_-^2\, (Y^-)^2
\,,
 \label{YLag}
\\[10pt]
&\Omega_+^2=\frac{\omega^2+2C^2}{4}\,,
\qquad
\Omega_-^2=\frac{\omega^2-2C^2}{4}\,,
\label{efffreq}
\end{align}
\label{tidLag}
\esub
which describes a \emph{time independent} (but still anisotropic) oscillator put into a \emph{constant} magnetic field. It has therefore natural time-translational symmetry,
$ U\to U+e $
(whereas rotations are still broken); transforming back
to the Brinkmann coordinates provides us precisely with the ``screw symmetry'' (\ref{screwtransf}) for the time-dependent system (\ref{tdLag}).

To get further insight, we note that the effective force in the $Y^-$ direction can be attractive, vanish or repulsive
depending on the sign of $\Omega_-^2$\,; the $Y^+$ direction is always repulsive, because $\Omega_+^2>0$.
The equations of motion which correspond to (\ref{tidLag}) are
\goodbreak
\besub\vspace{-7mm}
\begin{align}
&\ddot Y^+ - \omega \dot Y^- - {\Omega^2_+}\, Y^+ = 0,
\label{Omega-eq}
 \\[2pt]
&\ddot Y^- + \omega \dot Y^+ - {\Omega^2_-}\, Y^- =0.
\label{Omega+eq}
\end{align}\vspace{-3mm}
\label{Omegaeqs}
\esub
%
Particularly simple motions arise when \footnote{The system (\ref{Omegaeqs}) can actually be solved in full generality and without the assumption (\ref{correl}), but the formulae are substantially more complicated, see \cite{ZGH-Hill}.}
\beq
\omega^2=2C^2\qquad\text{\small when}\qquad \Omega_-=0,
\label{correl}
\eeq
so that the oscillator becomes maximally anisotropic.
The attractive term drops out from (\ref{tidLag}), leaving us  with a one-sided ``inverted'' (repulsive) oscillator in the $Y^+$ direction (plus an effective magnetic field). Simple solutions can be found readily in this case  \cite{ZGH-Hill}
\besub
\begin{align}
&Y^+ = \frac{\sqrt{2}\,c_1}{C} + c_2 \cos Ct +c_3 \sin Ct,
\\[2pt]
&Y^- = - c_1 t - \sqrt{2} c_2 \sin Ct + \sqrt{2} c_3 \cos Ct + c_4\,,
\end{align}
\label{specsol}
\esub
where $c_i,\, i=1,\dots 4$ are constants.  Then the inverse of (\ref{rotframe}) provides us with trajectories in Brinkmann coordinates.

\begin{figure}[h]
\begin{center}
\includegraphics[scale=.3]{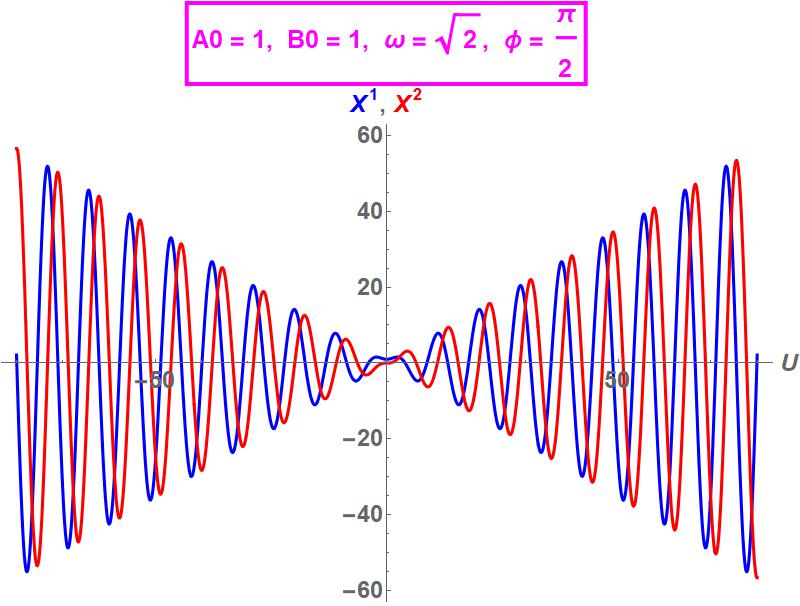}\quad
\includegraphics[scale=.25]{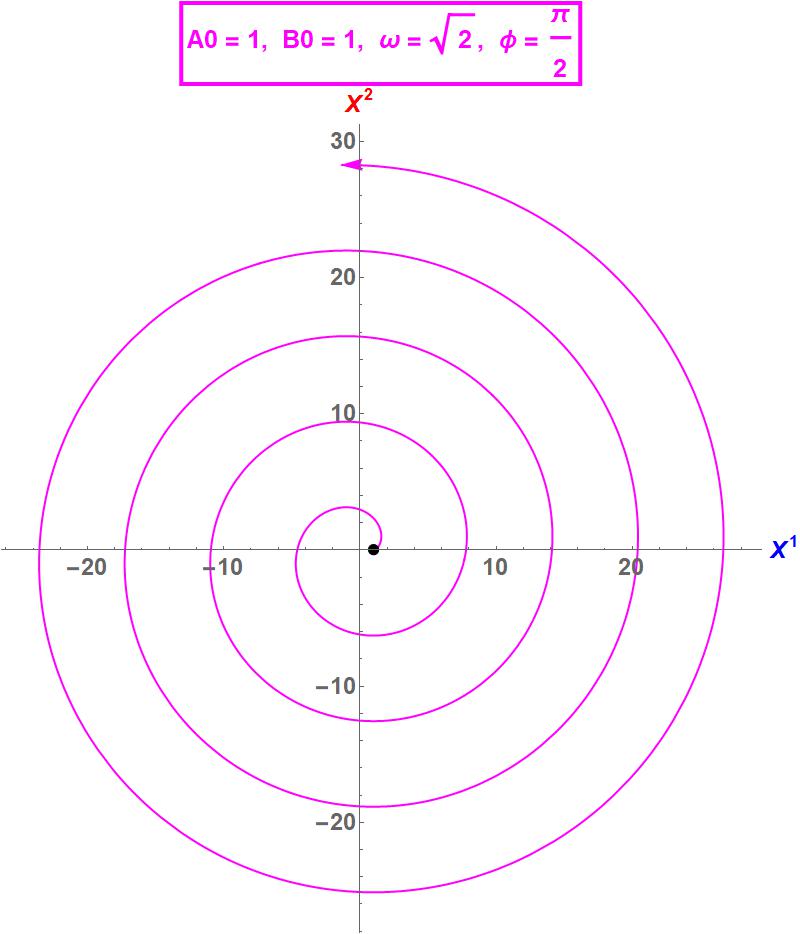}
\end{center}\vskip-5mm
\caption{\textit{\small When the amplitude and the frequency are correlated as in (\ref{correl}), a particle at rest at $U=0$ can, for appropriate initial conditions spiral outward. Here we took  $X^1(0)=0,\, X^2(0)=1,\,\dot{X}^1(0) = \dot{X}^2(0)=0$.
The increase of the radius is manifest.
}}
\label{specspiral}
\end{figure}

\begin{figure}[h]
\begin{center}
\includegraphics[scale=.27]{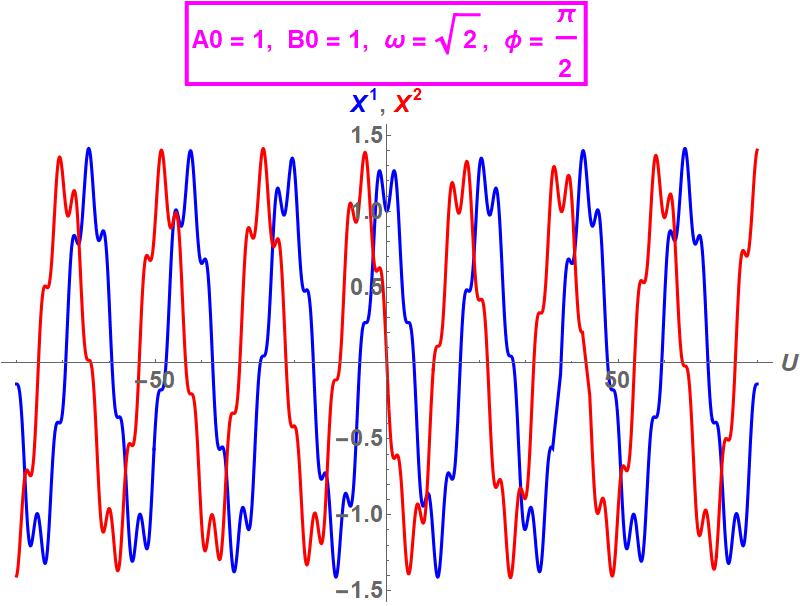}\;\;
\includegraphics[scale=.26]{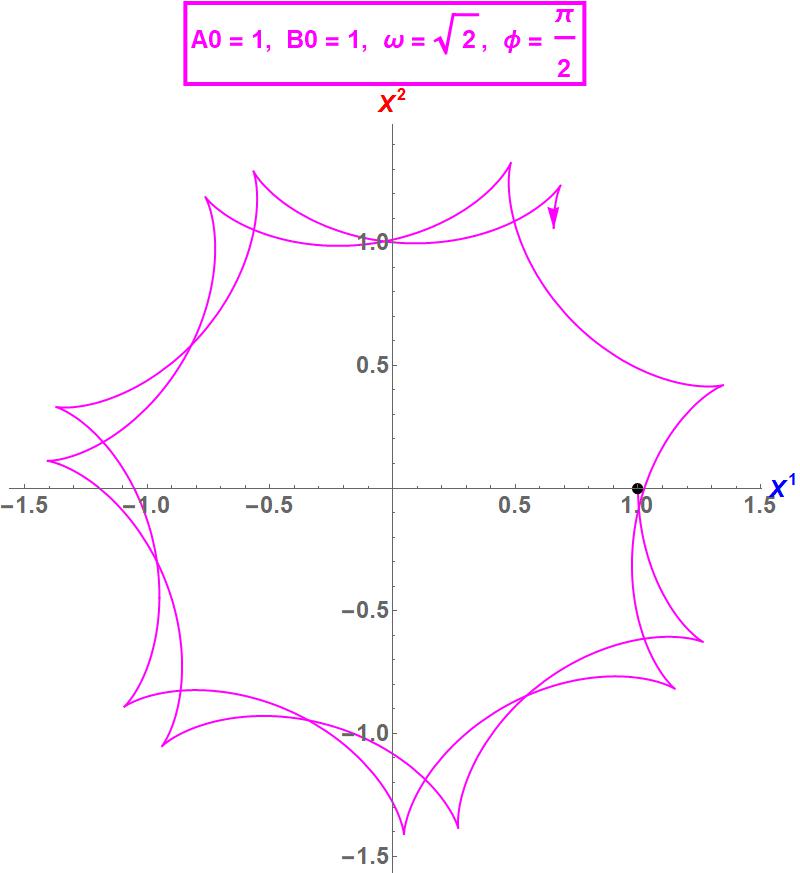}
\end{center}\vskip-8mm
\caption{\textit{\small  
When the amplitude and the frequency are correlated as in  (\ref{correl}), a particle at rest at $U=0$ can move periodically when appropriate initial conditions are chosen. Here we took $\blue{X^1(0)}=0,\, \red{X^2(0)}=1,\,\blue{\dot{X}^1(0)}=-1/\sqrt{2},\, \red{\dot{X}^2(0)}=0$.
}}
\label{specper}
\end{figure}

Choosing appropriate initial conditions we can arrange to get both  outwards-spiraling (Fig.\ref{specspiral}) or periodic (Fig.\ref{specper}) motion.
In the spiraling case the velocity diagram is essentially the same as for the trajectories and is therefore not reproduced here. 

\begin{figure}[h]\vspace{-3mm}
\begin{center}
\includegraphics[scale=.28]{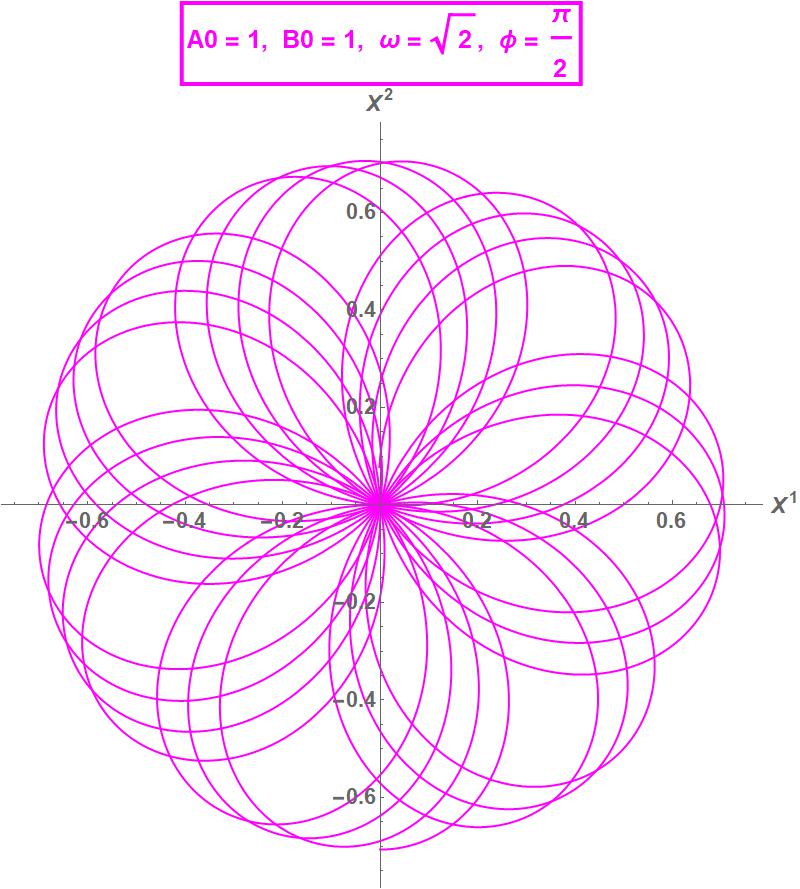}
\end{center}\vskip-8mm
\caption{\textit{\small Velocity diagram (hodograph) in  the periodic case of Fig.\ref{specper}.}}
\label{perhodograph}
\end{figure}

Periodic motions, (Fig.\ref{specper}), are obtained when the initial conditions satisfy
$\dot{X}^1(0)+(\omega/2)X^2(0)=0$.
The velocity diagram (called the \emph{hodograph}) shows a ``flower-like'' pattern, see Fig.\ref{perhodograph}.

In what follows, we shall explore in more depth
the symmetry enhancement  of polarised gravitational waves. We shall follow the treatment  \cite{Sippel} (see also \cite{Ehlers,exactsol,Siklos1,Siklos2}).
Row 10 of  Table II of \cite{Sippel} confirms that
the metric (\ref{genBrink}) has, in the general case, 5 Killing vectors, namely $\p_V,$ and
\beq
F(U)\,\p_1+E(U)\,\p_2-\big(X^1F'(U)+X^2E'(U)\big)\,\p_V,
\eeq
where
\beq
\left\{
\barraynb{lll}
F{\cA}+E{\cal A}_{\times} &=& F''
\\
F{\cal A}_{\times}-E\cA &=& E''
\earraynb\right.\,.\quad
\label{symeq}
\eeq
Equivalently, the symmetries are thus determined by
 $4$ first-order linear ODEs and
have therefore $4$ linearly independent solutions.
Inserting
the values read off from (\ref{PerProf}) yields,
\beq
\barraynb{cllll}
\;\;({C^2}/{2})\,\big( -F\cos(\omega U)&+&E\sin(\omega U)\big)&=& F''
\\[6pt]
({C^2}/{2})\,\big(\;\;\;F\sin(\omega U) &+&E\cos(\omega U)\big)&=& E''
\earraynb\;.
\label{G10}
\eeq
It is worth noting that these same equations could have been
obtained  by asking, as in \cite{GiPo}~:  \textit{``which time-dependent translations,
\beq
\bX \to \bX+\ba(U),\qquad
\ba(t) =\barray{c}E(U)\\ F(U)\earray\,,
\eeq
 preserve the equations of motion (\ref{Xeqmot})~?''}
Comparing (\ref{G10}) with (\ref{Xeqmot}) leads to the
remarkable conclusion that \emph{the symmetry equations, (\ref{G10}), are, after rescaling,
identical to the equations of motion satisfied by the coordinates}. Having solved the latter we had also determined the isometries~: they form a central extension of the Newton-Hooke group with broken rotations \cite{ZGH-Hill}.


Noether's theorem applied to the geodesic Lagrangian  (\ref{geoLag}) associates a constant of the motion along $X^{\mu}(\sigma)$,
\beq
{\cal E}[K^{\mu}]= K^{\mu} g_{\mu \nu}\frac{dX^{\nu}}{d\sigma}\,,
\label{Noetherquant}
\eeq
to any Killing vector field $K^\mu$.
For those 5 standard generators this has been done, e.g., in \cite{ZGH-Hill}; for ${\partial}_{V}$ in particular we get $\dot U = \const$
which confirms that $U$ is indeed an affine parameter.
For the extra ``screw symmetry'' (\ref{screwtransf}) we find in turn an extra Killing vector,
\beq
K^{\mu} \partial_{\mu} = \partial_U  + \frac{\omega}{2} (X^1 \partial_2 - X^2\partial_1)\,,
\eeq
to which Noether associates
\beq
{\cal E}= H_{ij}X^iX^j + \dot{V} + \frac{\omega}{2} \bigl(X^1 \dot X^2 - X^2\dot X^1\bigr)
\eeq
as we could check numerically for the non-trivial periodic solution depicted in Fig.\ref{specper}.

The Lie  algebra of Killing vectors can also be determined. If
$F,E=(F_1,E_1), (F_2,E_2) $ respectively then the
Lie bracket of  two vector fields  is of the form
$$
-\bigl(F_1 F_2^{\prime} - F_2 F_1^{\prime}  + E_1 E_2^{\prime} - E_2 E_1^{\prime} \bigr) \,\p_V\,.
$$
Now
\beq
(F_1 F_2^{\prime} - F_2 F_1^{\prime}  + E_1 E_2^{\prime} - E_2 E_1^{\prime}\bigr)^{\prime}
= F_1 F_2^{\prime \prime} - F_2 F_1^{\prime \prime}  +   E_1 E_2^{\prime \prime} -
E_2 E_1^{\prime \prime} \,.
\eeq
 Then using (\ref{symeq}) we deduce that the Lie bracket
of every 6 pairs of such vector fields is a constant multiple
of the  central element $\p_V$. Thus the five dimensional
symmetry group is a central extension of a four dimensional Abelian group.

We now turn to the structure of the enhanced algebra.
We compute the bracket
\begin{eqnarray}
&&\Big[\p_U+ \omega X^1 \p_2- X^2 \p_1),
F\p_1+E\p_2+(X^1 F^{\prime} +X^2 E^{\prime})\p_V\Big]\\
&=&
F ^{\prime} \p_1+E^{\prime} \p_2+(X^1 F''+X^2 E'' )
\p_V+\omega (X^1 F^\prime -X^1 E^\prime )\p_V.
\end{eqnarray}

The subalgebra could be 2,3, or 4 dimensional.
If three dimensional, it could be one of the Bianchi algebras.
This case has been investigated  in \cite{Siklos1}.
Inspection  reveals that it is not our case.
Somewhat earlier \cite{Ehlers}
the issue of extra symmetries had also been considered.
For a six dimensional group two cases are listed,  ours, and the case considered in \cite{Siklos1,Siklos2}.
It is claimed that there is a four-dimensional simply
transitive subgroup but no details are given.

\section{Conclusion}

Our investigations confirm that under the influence of a polarized sandwich wave (imitated here by an oscillating Gaussian profile)
test particles originally at rest move, in the (approximately flat) before and after zones, along straight lines with constant velocity --- just as they do in the linearly polarized case and as expected theoretically. However motion in the inside zone is a rather complicated combination of oscillations and precessions.

Shrinking the Gaussian to a Dirac delta by letting the thickness parameter go to infinity, much of the complications (including the precessions) disappear cf. \cite{PodVes} and the behaviour in an impulsive wave \cite{ImpMemory,PodSB,PodVes} is recovered.

Widening the Gaussian by letting the thickness parameter go to zero yields instead, after a suitable rescaling, periodic GWs, which ``only have an inside zone".
The precession as well as the ``breathing" of the ``dented'' trajectories become manifest.
The model exhibits an intriguing additional ``screw symmetry'' noticed before \cite{Sippel,exactsol,Carroll4GW}.

\kikezd{Note added}. 
 Bondi \cite{Bondi57}, in his note establishing the physical reality of gravitational waves and putting an end to the ongoing controversy  writes:  
``\textit{\narrower  
Consider a set of test particles at rest
[\dots] before the arrival of the wave. Then, after the passage of the wave, the particle that was at rest will have} [a non-trivial] \textit{velocity $4$-vector \dots}''

Finally, the recent papers  \cite{FaSu} and 
\cite{Ilderton} contribute to our understanding of the memory effect.

\begin{acknowledgments}
 We thank L. Blanchet, M. Faber, B. Kocsis, P. Kosinski, A. Lasenby, S.T.C. Siklos
  for their interest advice and correspondence. PH is grateful to the \emph{Institute of Modern Physics} of the Chinese Academy of Sciences in Lanzhou for hospitality. Support by the National Natural Science Foundation of China (Grant No. 11575254) is acknowledged.
\end{acknowledgments}
\goodbreak


\end{document}